# Stacking-Dependent Electronic Properties in GaSe/GaTe Heterobilayers: A First-Principles Study

Hsin-Yi Liu and Jhao-Ying Wu*

Department of Energy and Refrigerating Air-Conditioning Engineering

, National Kaohsiung University of Science and Technology, Kaohsiung 811213, Taiwan

## Abstract

In this study, we use first-principles calculations to investigate the stacking-dependent electronic properties of GaSe/GaTe van der Waals heterobilayers. By analyzing five representative stacking configurations–AA, AA′, A′C, A′B, and AB–we show that interlayer atomic registry affects orbital hybridization and interfacial interactions, leading to distinct electronic structures and stabilities. Projected density of states analyses reveal valence and conduction band edges arise from orbitals localized in different layers, confirming a type-II band alignment that facilitates spatial charge separation. Orbital contributions and spectral features vary with stacking, reflecting how interlayer coupling modulates hybridization and electronic behavior. This study provides atomic-level insights for designing and optimizing layered heterostructures in nanoelectronic devices.

# 1 Introduction

Two-dimensional (2D) materials have attracted considerable attention due to their unique physical properties and potential for integration into next-generation electronic and optoelectronic devices [1-4]. Their structural, electronic, and interlayer characteristics can be precisely tuned by external perturbations and stacking configurations, offering opportunities for tailored device functionalities [5-6]. Among various 2D material systems, group III-VI metal monochalcogenides, such as gallium selenide (GaSe) and gallium telluride (GaTe), stand out due to their intrinsic semiconducting nature, layer-dependent bandgaps [7-10], strong in-plane anisotropy [11, 12], and environmental stability [7, 13, 14]. These properties provide practical advantages over graphene, which lacks a native bandgap [15], and black phosphorus, which suffers from rapid degradation in ambient conditions [16, 17]. As such, GaSe and GaTe offer promising platforms for both fundamental studies and real-world applications.

Monolayer GaSe and GaTe adopt a hexagonal structure composed of a four-atom-thick Ga–X–Ga–X (X = Se, Te) stacking, where each Ga atom forms trigonal prismatic bonds with adjacent chalcogen atoms. This configuration results in an indirect bandgap in the bulk phase, which evolves toward a more direct-like character as the number of layers decreases [18-20]. However, even at the monolayer limit, these materials retain an indirect bandgap due to the momentum mismatch between the valence and conduction band edges [7, 21]. This indirect nature not only suppresses radiative recombination efficiency but also weakens optical absorption near the band edge, thereby limiting their potential in optoelectronic applications such as light emission and photodetection [11, 22].

To overcome such limitations and achieve better control over the electronic structure,

vertical heterostructures comprising dissimilar 2D layers have emerged as a powerful strategy [2, 23, 24]. These van der Waals (vdW) heterostructures offer new functionalities by engineering interlayer coupling, interface chemistry, and orbital hybridization [25, 26]. In this context, GaSe/GaTe heterobilayers serve as an ideal platform for investigating how lattice mismatch, interfacial strain, and symmetry breaking influence the resulting electronic structure [27, 28]. The structural and chemical complementarity of GaSe and GaTe allows for diverse stacking configurations, each potentially leading to distinct electronic behaviors through modified atomic registries and orbital overlap [29].

Previous studies on analogous systems, such as $MoS_2$/$WS_2$ and $WSe_2$/$MoSe_2$ heterostructures, have shown that stacking arrangements can induce substantial shifts in band alignments and even transitions in the bandgap character [30, 31]. However, a systematic theoretical investigation focusing on the electronic consequences of stacking in GaSe/GaTe heterostructures remains lacking. Understanding the microscopic mechanisms that govern stacking-dependent modulation of interlayer interactions, charge transfer, and orbital hybridization is critical for the rational design of heterostructures with desirable electronic functionalities.

In this work, we conduct a comprehensive first-principles study to elucidate the pivotal role of stacking configurations in modulating the electronic properties of GaSe/GaTe heterobilayers. Our results suggest that variations in stacking order induce substantial changes in orbital hybridization, interlayer interactions, and spatial charge distribution. These effects collectively influence the band alignment and bonding nature of the system. Notably, projected density of states (PDOS) analyses highlight the orbital origins of these stacking-dependent phenomena, illustrating how distinct atomic contributions dictate the tunability of the electronic structure. This orbital-level understanding establishes a direct correlation between stacking geometry and electronic modulation, offering valuable insights into the rational design of group III-VI van der Waals heterostructures for next-generation nanoelectronic devices.

## 2 Methods

First-principles calculations based on density functional theory (DFT) were performed using the VASP package [32]. The projector-augmented wave (PAW) method [33, 34] was used to describe the electron-ion interactions. Exchange-correlation effects were treated with the Perdew-Burke-Ernzerhof (PBE) generalized gradient approximation [35]. A plane-wave cutoff energy of 500 eV was employed. To prevent spurious interactions between periodic images, a vacuum spacing of 20 Å was applied along the direction perpendicular to the layers. Van der Waals (vdW) interactions were included via Grimme's semiempirical DFT-D3 correction with Becke-Johnson damping [36]. Structural optimizations were carried out until the Hellmann-Feynman forces were below 0.01 eV/Å and the total energy change was less than $10^{-5}$ eV. For geometry optimization and electronic structure calculations [37], the Brillouin zone was sampled using a 100×100×1 k-point mesh centered at the $\Gamma$ point. Projected density of states (PDOS) calculations employed a denser 150×150×1 k-point grid.

# 3 Geometric Structures

The geometric configurations of monolayer GaSe and GaTe are displayed in Figs. 1(a) and 1(b), respectively. Both materials feature a buckled atomic structure along the $z$-axis, as shown in the upper panels. Within each layer, Ga and chalcogen atoms (Se or Te) are arranged in a trigonal prismatic coordination, forming a quasi-two-dimensional honeycomb lattice in the $xy$-plane, as illustrated in the lower panels. The buckling height $\Delta h$, defined as the vertical displacement between Ga and chalcogen atoms and indicated in Fig. 1(a), plays a crucial role in modulating orbital hybridization and enhancing out-of-plane polarizability.

For the heterobilayer system, five representative stacking configurations are considered, as shown in Figs. 1(c)-1(g): AA, AA′, A′C, A′B, and AB. These structures differ in the relative lateral positioning of the top GaSe layer and the bottom GaTe layer. In the AA stacking (Fig. 1(c)), the Ga and chalcogen atoms of the upper and lower layers are aligned directly above one another. The AA′ stacking (Fig. 1(d)) corresponds to a configuration where Ga atoms in the top layer are aligned with chalcogen atoms in the bottom layer and vice versa. In the A′C and A′B configurations (Figs. 1(e) and 1(f)), each atom in one layer sits approximately at the center of a hexagon in the other layer. Finally, the AB stacking (Fig. 1(g)) involves a partial shift of the sublattices, commonly observed in other layered materials.

The stacking configuration plays a pivotal role in determining the interlayer distance, atomic registry, and total energy. These structural variations directly influence orbital overlap and interlayer interactions, which are crucial for modulating the electronic band structure. The optimized geometric parameters for all systems are summarized in Table 1. The interlayer distance $D$, as illustrated in Fig. 1(c), ranges from 3.917 to 3.985 Å depending on the stacking pattern. These values are larger than those reported for bilayer graphene (3.26–3.53 Å), silicene (2.43–2.72 Å), germanene (2.428–2.53 Å) [38, 39], and GaS (3.746–3.822 Å) [40]. Among the five configurations, A'C exhibits the largest interlayer spacing, while A'B shows the smallest. Additionally, the AA' structure is the most energetically favorable, as evidenced by its lowest total energy per atom ($E_0$).

The interlayer coupling significantly influences the bond lengths and buckling characteristics of the constituent layers. In the GaSe layer of the GaTe/GaSe heterostructure, both the Ga–Ga and Ga–Se bond lengths are slightly elongated compared to those in monolayer GaSe, accompanied by a reduction in the buckling height across all stacking configurations. In contrast, the GaTe layer exhibits shorter Ga–Ga and Ga–Te bond lengths relative to monolayer GaTe, along with an increased buckling height. These structural changes highlight the complex interplay between interlayer interactions and lattice mismatch, which are crucial for determining the overall properties of the heterostructure.

# 4  Energy Bands

A comprehensive understanding of the electronic properties of monolayer GaTe and GaSe is essential, as they serve as the fundamental building blocks for constructing GaSe/GaTe heterostructures with tunable electronic behaviors. As shown in Fig. 1, both monolayers exhibit indirect band gaps, with the valence band maximum (VBM) and conduction band minimum (CBM) located at different $k$-points. For GaTe (Fig. 1(a)), the VBM lies between the $\Gamma$ and M points, while the CBM appears between the M and K points. In contrast, GaSe (Fig. 1(b)) has its CBM positioned at the $\Gamma$ point. These differences originate from variations in atomic composition, which modulate orbital hybridization, lattice constants, and intralayer interactions, thereby reshaping the conduction band dispersion.

An analysis of the valence band structure shows that the VBM in both GaTe and GaSe originates from quasi-$\pi$ bands spanning the energy range from $-0.7$ to $-2$ eV. In GaTe, these bands display significant dispersion, reaching a minimum at the K point and exhibiting strong hybridization with $\sigma$ bands, as evidenced by band anticrossings near the M and K points. In GaSe, the quasi-$\pi$ bands form a composite parabolic profile along the $M \to K \to \Gamma$ path, with a distinct saddle point at M where $E^v = -2$ eV.

Examination of the $\sigma$ bands further reveals that, in GaTe, they emerge from the $\Gamma$ point at approximately $E^v = -0.8$ eV and extend down to $-4.2$ eV at K, while in GaSe, they span a slightly broader range from $-1.2$ to $-4.5$ eV. In both materials, multiple band anticrossings between $\pi$ and $\sigma$ states are observed, indicating strong wavefunction hybridization and the breakdown of pure orbital character. Additionally, deep $\sigma$ bands anticross with Ga $s$ states near $E^v \approx -4$ eV, further confirming the robust $sp^3$ hybridization characteristic of these layered systems. The relatively elevated energy position of these $sp^3$ features in GaTe and GaSe–compared to the deeper $-10$ eV levels in monolayer graphene– underscores the combined effects of their puckered lattice geometry and the presence of heavier chalcogen atoms.

In GaTe/GaSe heterobilayers (Fig. 2(b)-(f)), each monolayer-derived band splits into bonding and antibonding subbands due to interlayer orbital hybridization, effectively doubling the number of subbands. Unlike homobilayers (e.g., bilayer GaS), where inversion symmetry preserves certain subband degeneracies at high-symmetry points, the GaTe/GaSe heterobilayer lacks inversion and mirror symmetries due to the chemical and structural asymmetry between the constituent layers. This results in the lifting of all subband de-generacies across the Brillouin zone. The degree of splitting is amplified by the contrast in onsite energies, orbital compositions, and asymmetric interlayer coupling between GaTe and GaSe interactions.

The stacking arrangement plays a pivotal role in determining the extent of orbital overlap between Ga and chalcogen atoms, thereby shaping the subband dispersions and overall electronic topology. Despite variations in stacking, all heterobilayer configurations consistently display a direct band gap at the $\Gamma$ point, with both the CBM and VBM anchored at this location–marking a distinct contrast to the indirect gaps found in the monolayers. The direct band gap in GaTe/GaSe offers clear advantages for optoelectronic devices. Direct-gap materials exhibit superior light-matter interaction strength, enabling efficient optical absorption and radiative recombination. Consequently, the GaTe/GaSe heterobilayer is particularly well-suited for light-emitting diodes, lasers, and photodetectors.

Among the stacking configurations considered, AA and A'C exhibit the smallest energy gaps (Table 1), which may be attributed to their relatively favorable vertical atomic alignments. This arrangement appears to enhance interlayer orbital interactions, potentially contributing to the observed modifications in the electronic band structure. In particular, saddle points are identified at the M point, with their number and degeneracy varying depending on the stacking configuration. Moreover, changes in the band curvature near the extrema are observed, which affect the effective masses of charge carriers. For instance, reduced effective masses at the $\Gamma$ point suggest that certain stacking configurations could be advantageous for charge transport.

# 5  Spatial Charge Density

To provide a representative comparison among the five stacking configurations (AA, AA′, A′C, A′B, and AB), we focus on AA′ and A′C stackings in the following analysis of spatial charge density and density of states (DOS). The AA′ configuration exhibits the lowest total energy and the widest bandgap, indicating the most stable and insulating nature. In contrast, the A′C stacking shows the highest total energy and the smallest bandgap, reflecting the least stable and most conductive behavior. Therefore, these two cases serve as limiting examples to highlight the influence of stacking on electronic characteristics.

Spatial charge density is a crucial probe for investigating hybridization between multiple and single orbitals, which plays a pivotal role in determining the electronic structure of layered materials. The total charge density distributions ($\rho$) for the AA′ and A′C stackings, shown in Figs. 4(a) and 4(b), provide important insights into the nature of interatomic bonding across various cross-sections. These visualizations highlight the Ga–Ga, Ga–Se, and Ga–Te interactions and illustrate how different stacking configurations modulate the bonding characteristics of GaTe/GaSe heterostructures.

For Ga–Ga bonds, as shown in the cross-sectional views of Ga atoms, the $\sigma$ bonding (yellow regions) primarily originates from $sp^3$ hybridization involving Ga-(4s, $4p_x$, $4p_y$, $4p_z$) orbitals. This hybridization reflects the directional covalent bonding typical of the puckered lattice structures in GaTe and GaSe. Additionally, weak $\pi$ bonding (green regions) is observed, likely arising from partial side-by-side overlap of unhybridized or partially hybridized $p$ orbitals. These $\pi$ interactions may result from non-ideal hybridization or local symmetry distortions. The resulting bond angles deviate from the ideal tetrahedral geometry, which influences both bond strength and charge distribution.

In contrast, the Ga–Se and Ga–Te bonds exhibit stronger $\sigma$ bonding (red regions) than the Ga–Ga bonds, which can be attributed to an $sp^2$-like hybridization between Ga (4s, $4p_x$, $4p_y$) and Se (4s, $4p_x$, $4p_y$) or Te (5s, $5p_x$, $5p_y$) orbitals. This type of hybridization, associated with smaller bond angles compared to $sp^3$, allows for more efficient orbital overlap, thereby enhancing bond strength. Furthermore, the cross-sectional views of Se and Te atoms display isotropic regions of high charge density, primarily arising from their $s$ orbitals. This isotropy results from the spherical symmetry of the $s$ orbitals, whose non-directional nature limits their involvement in covalent bonding. The accumulation of charge density in these orbitals contributes to structural stabilization by providing additional charge screening and facilitating localized charge redistribution around the anion sites.

Notably, the charge density analysis reveals stacking-dependent variations in bond strength. In the AA′ stacking, Te atoms in the GaTe layer are positioned directly above Ga atoms in the GaSe layer, leading to strong interlayer electrostatic attraction and orbital hybridization between Te and Ga across the layers. This enhanced vertical interaction induces local geometric relaxation, which shortens the intralayer Ga–Te bond length and strengthens the Ga–Te bonds in the GaTe layer. Conversely, in the A′C stacking, Se atoms in the GaSe layer are more closely aligned with Ga atoms in the GaTe layer, intensifying interlayer interactions on the GaSe side. As a result, the intralayer Ga–Se bonds are strengthened due to a similar bond-shortening mechanism. Therefore, stacking-induced variations in interlayer atomic alignment selectively reinforce intralayer bonding in the respective layers.

This stacking-dependent behavior underscores the tunable nature of interlayer interactions in van der Waals heterostructures, offering a mechanism to fine-tune electronic properties via stacking order. The differential bond strengths in each configuration signif-icantly affect the system's stability, electronic structure, and potential for applications in optoelectronic devices. Moreover, variations in charge distribution and bonding characteristics among different stacking arrangements are expected to influence charge transport, photonic response, and mechanical properties, thereby playing a key role in optimizing these materials for specific functionalities.

# 6  Density of States

To better understand the electronic structure and interlayer coupling in GaSe/GaTe heterobilayers, we analyze the density of states (DOS) and projected DOS (PDOS). These spectra capture the band dispersions—including quasi-$\pi$ and $\sigma$ bands, hybridizations, and anticrossing features—and directly relate to experimental probes such as scanning tunneling spectroscopy (STS). Notably, shoulders, symmetric peaks, and asymmetric subpeaks in the DOS correspond to van Hove singularities arising from parabolic band edges, saddle points, and flat bands, respectively. For example, in Fig. 5(a) (AA′ stacking), the shoulder at $E = -0.4$ eV originates from the VBM, while the peaks at $E = -1.2$ and $-2.5$ eV arise from flat bands along the $\Gamma \to M$ and $M \to K$ directions, respectively.

These spectral features reveal the underlying orbital contributions and band characteristics of the individual layers. To elucidate this behavior, we examine the layer-resolved PDOS, focusing on the orbital character and energy alignment. In both AA′ and A′C stackings (Figs. 5 and 6), the GaTe layer's $p_{x,y}$ orbitals dominate the VBM due to the higher on-site energy and larger spatial extent of Te $5p$ orbitals, which facilitate stronger hybridization compared to the Se $4p$ orbitals in GaSe. This dominance persists despite stacking-induced changes in local bonding environments, such as bond shortening and strengthening, as observed in the charge density plots. Therefore, the GaTe layer's contribution to the VBM remains robust across different stacking configurations.

Beyond the dominant $p$ orbitals, Ga $s$ orbitals also play a significant role in orbital hybridization. In both AA′ and A′C stackings (Figs. 5 and 6), Ga $s$ orbitals contribute more prominently than those of Se and Te due to their greater delocalization, which promotes $sp^3$-like hybridization with neighboring $p$ orbitals near the VBM. In contrast, the more localized $s$ states of Se and Te lie deeper in energy and mainly contribute to low-energy spectral features. A distinct $sp^3$ signature appears at $E = -3.9$ eV for Ga-1 atoms and at $E = -4.5$ eV for Ga-2 atoms (Figs. 5(b)(d) and 6(b)(d)), indicating that interlayer coupling alters the alignment of $s$-derived states across layers.

The stacking sequence significantly influences the DOS, modulating spectral intensity, energy positions, and peak multiplicity. Compared to the AA′ configuration (Fig. 5), the A′C stacking (Fig. 6) exhibits more pronounced subpeaks, mainly originating from Se and Te contributions. Se dominates the energy range from $E = -3.1$ to $-2.5$ eV, while Te contributes more significantly from $E = -2.5$ to $-1.9$ eV. Although spatial charge density analysis indicates stronger Se–Ga bonding in the A′C stacking, the PDOS reveals that Te–Ga hybridization near the valence band maximum is more pronounced. This enhanced Te–Ga orbital interaction leads to an upward shift of the GaTe $p_{x,y}$ orbitals and consequently reduces the band gap. This PDOS analysis highlights the interplay among atomic orbitals, stacking order, and interlayer interactions in GaSe/GaTe heterobilayers, providing a foundation for orbital engineering and band structure tuning in future optoelectronic and spintronic devices.

It is worth noting that the band-edge states exhibit clear spatial and orbital separation:

the VBM is predominantly derived from in-plane Te $p_{x,y}$ orbitals in the GaTe layer, supporting hole transport within the layer. In contrast, the CBM primarily originates from Ga $s$ orbitals in the GaSe layer, forming extended Bloch states that facilitate electron mobility. This type-II band alignment leads to efficient charge separation, with electrons and holes spatially confined in different layers. Such orbital and spatial decoupling reduces electron-hole recombination and enhances the performance of optoelectronic devices such as photodetectors, solar cells, and LEDs.

# 7 Conclusions

In summary, we have systematically investigated the electronic properties of GaSe/GaTe heterobilayers with various stacking configurations. Our results demonstrate that stacking order plays a crucial role in modulating the electronic structure, with the AA′ stacking exhibiting the widest energy gap and highest structural stability, while the A′C configuration shows a noticeable reduction in the energy gap, which may be associated with enhanced interlayer hybridization. Analyses of spatial charge density and projected density of states further suggest that stacking-dependent local geometric relaxations and orbital interactions are key factors in tuning the electronic behavior. Specifically, the heterobilayer consistently exhibits a type-II band alignment, where the valence band maximum is predominantly derived from Te $p_{x,y}$ orbitals in the GaTe layer, and the conduction band minimum mainly originates from Ga $s$ orbitals in the GaSe layer, facilitating effective interlayer charge separation. Overall, the present work deepens the understanding of interlayer coupling and offers practical guidance for tailoring the electronic properties of van der Waals layered semiconductors.

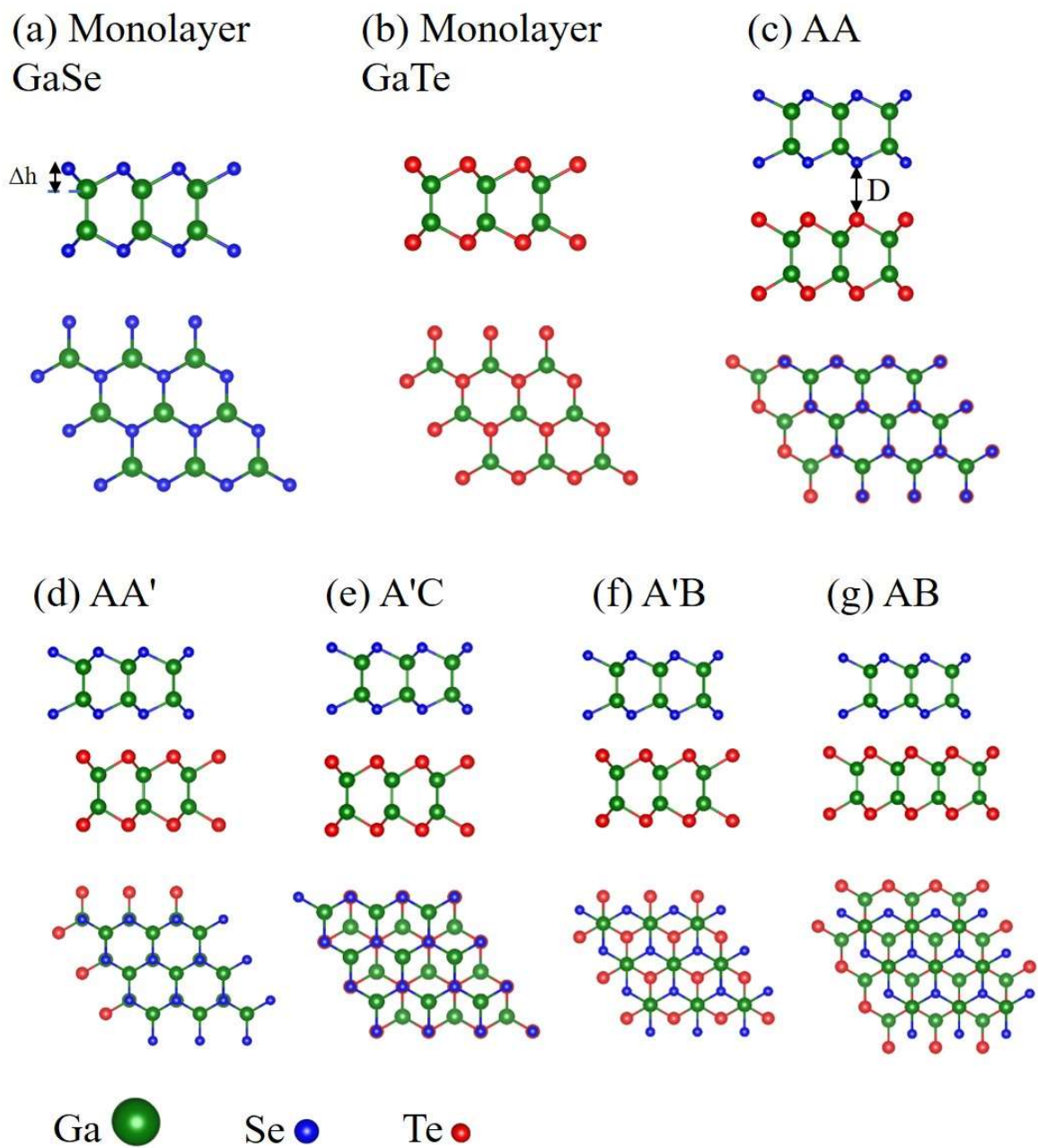

Figure 1: Side (upper panels) and top (lower panels) views of (a) monolayer GaSe, (b) monolayer GaTe, and (c)–(g) various stacking configurations of GaSe/GaTe heterobilayers, including (c) AA, (d) AA′, (e) A′C, (f) A′B, and (g) AB. Green, blue, and red spheres represent Ga, Se, and Te atoms, respectively. $\Delta h$ and $D$ denote the buckling height and interlayer distance, respectively.

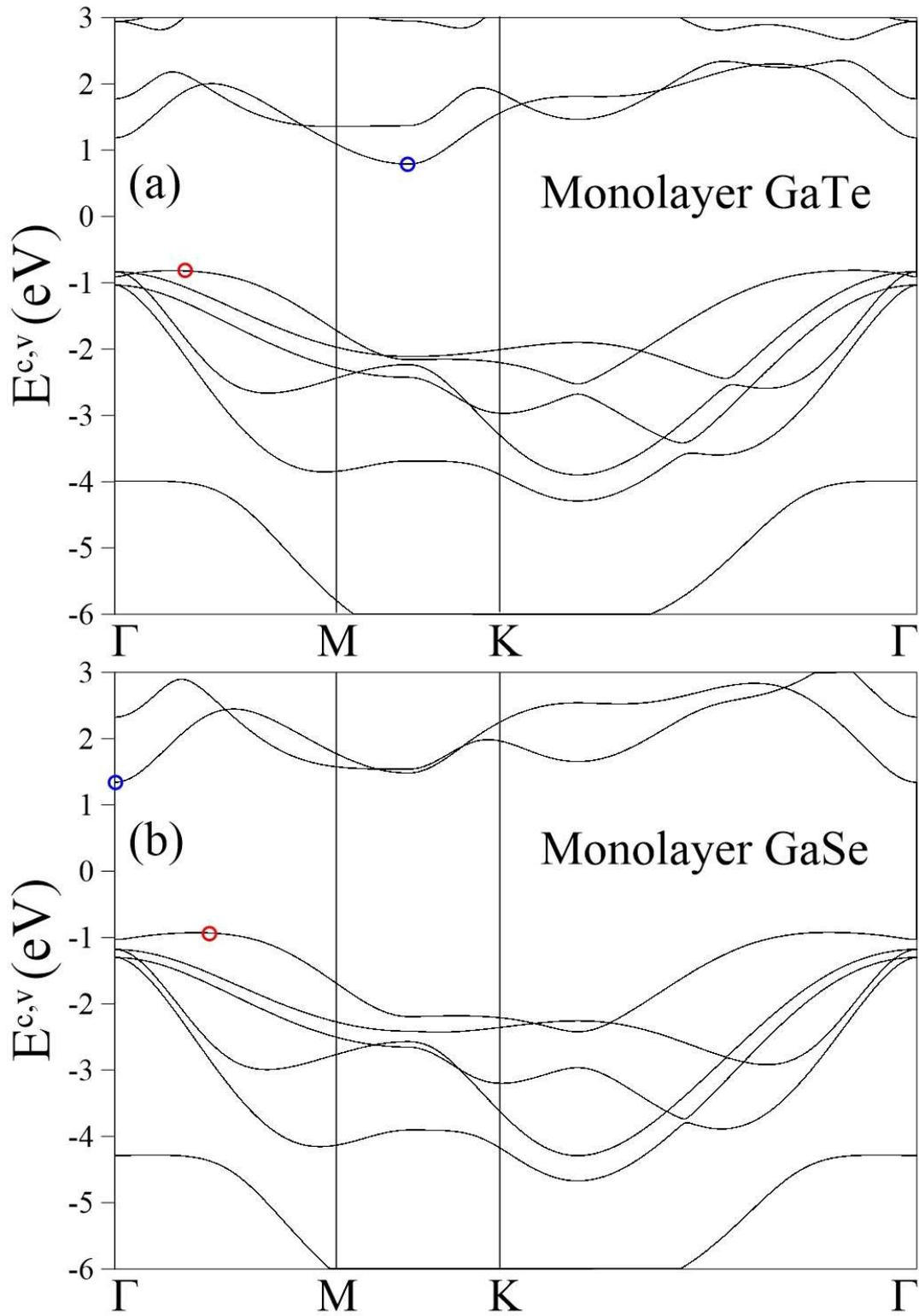

Figure 2: Electronic band structures of (a) monolayer GaTe and (b) monolayer GaSe. The conduction band minimum (CBM) and valence band maximum (VBM) are marked by blue and red circles, respectively. The Fermi level is set to zero energy.

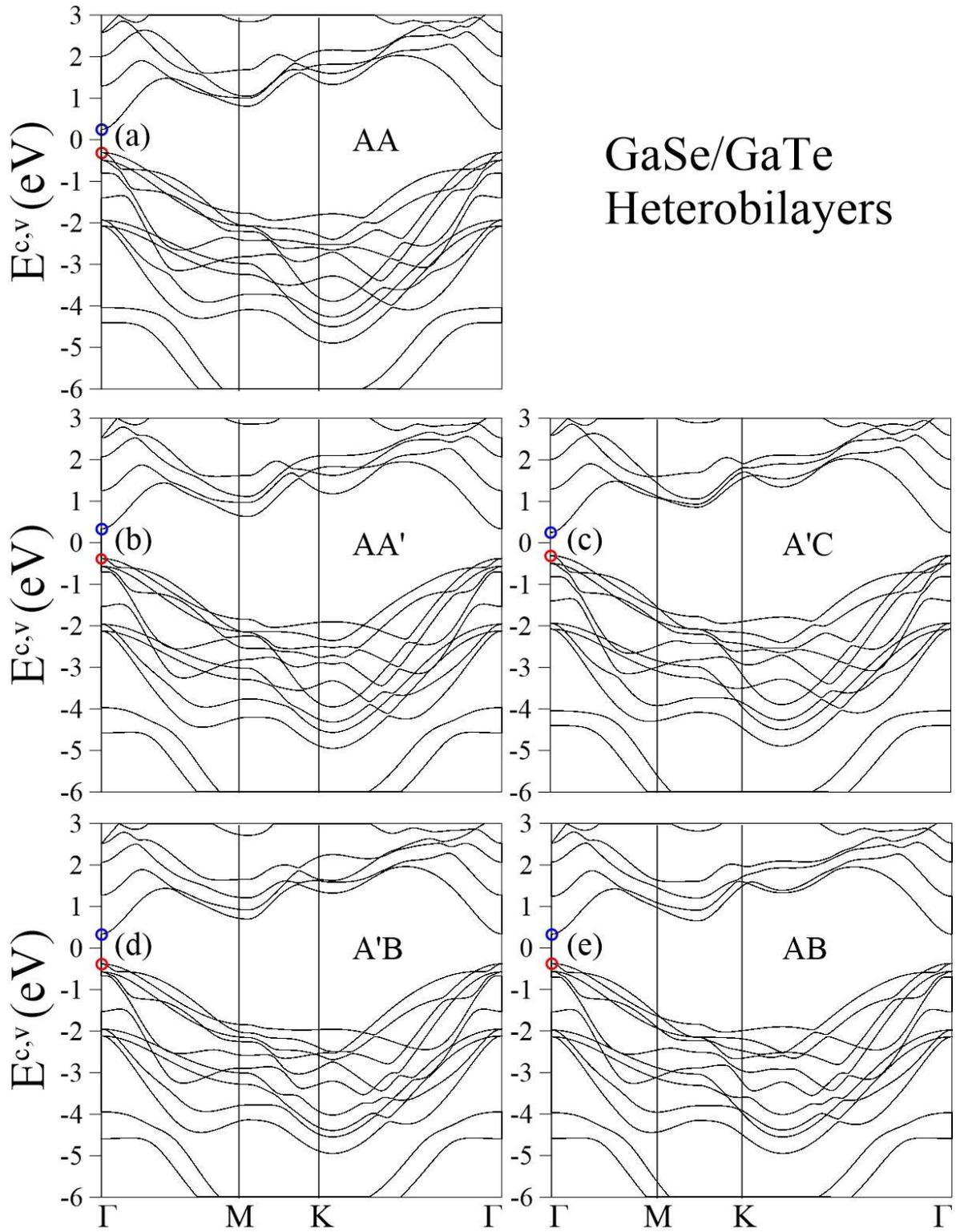

Figure 3: Electronic band structures of GaSe/GaTe heterobilayers with different stacking configurations: (a) AA, (b) AA′, (c) A′C, (d) A′B, and (e) AB. The conduction band minimum (CBM) and valence band maximum (VBM) are highlighted by blue and red circles, respectively. The Fermi level is set to zero energy.

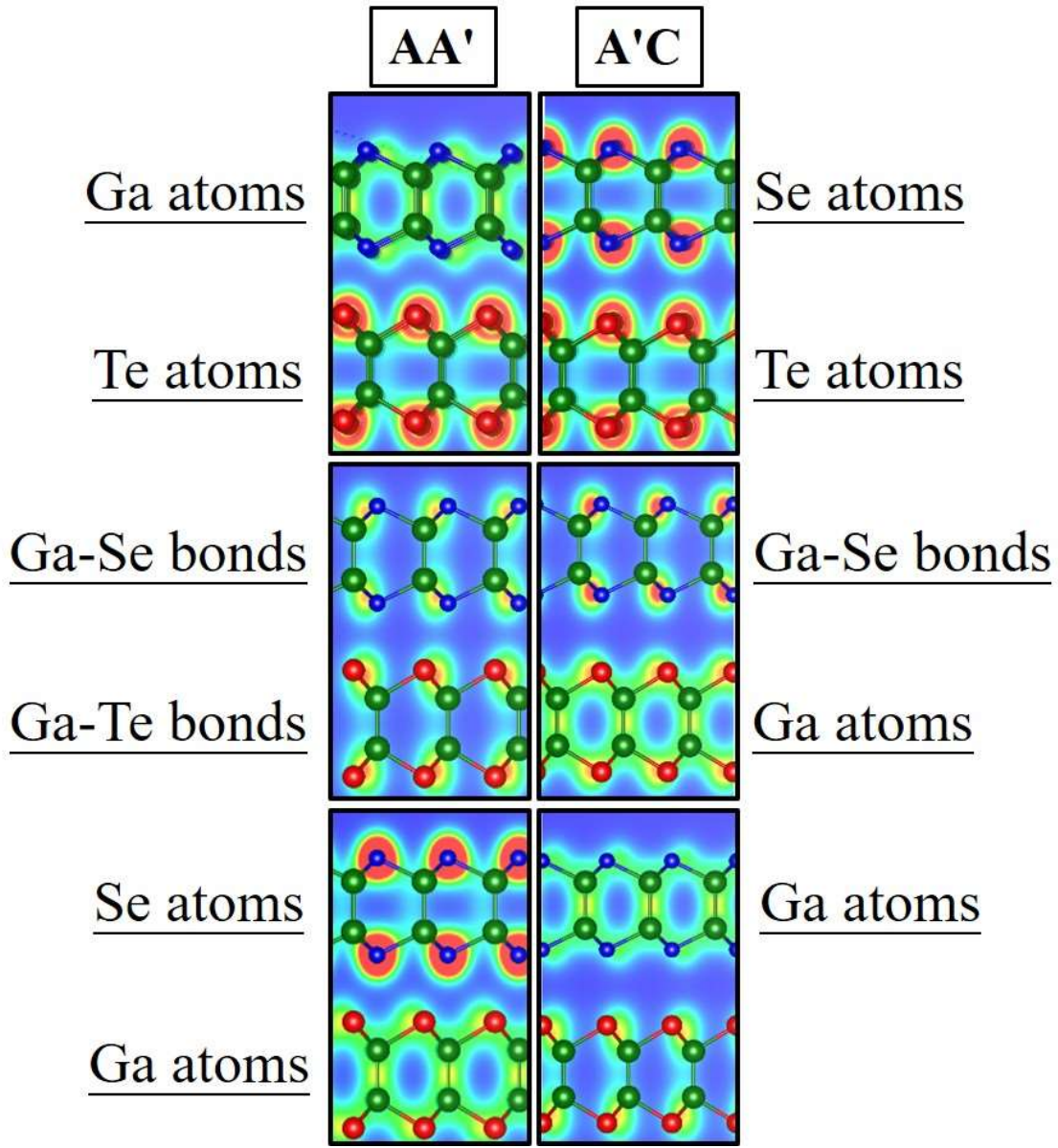

Figure 4: Spatial charge densities ($\rho$) of the (a) AA′- and (b) A′C-stacked GaSe/GaTe heterobilayers. Each panel presents a cross-sectional view of the charge distribution at specific atomic planes or bonding regions, as labeled on the left and right sides.

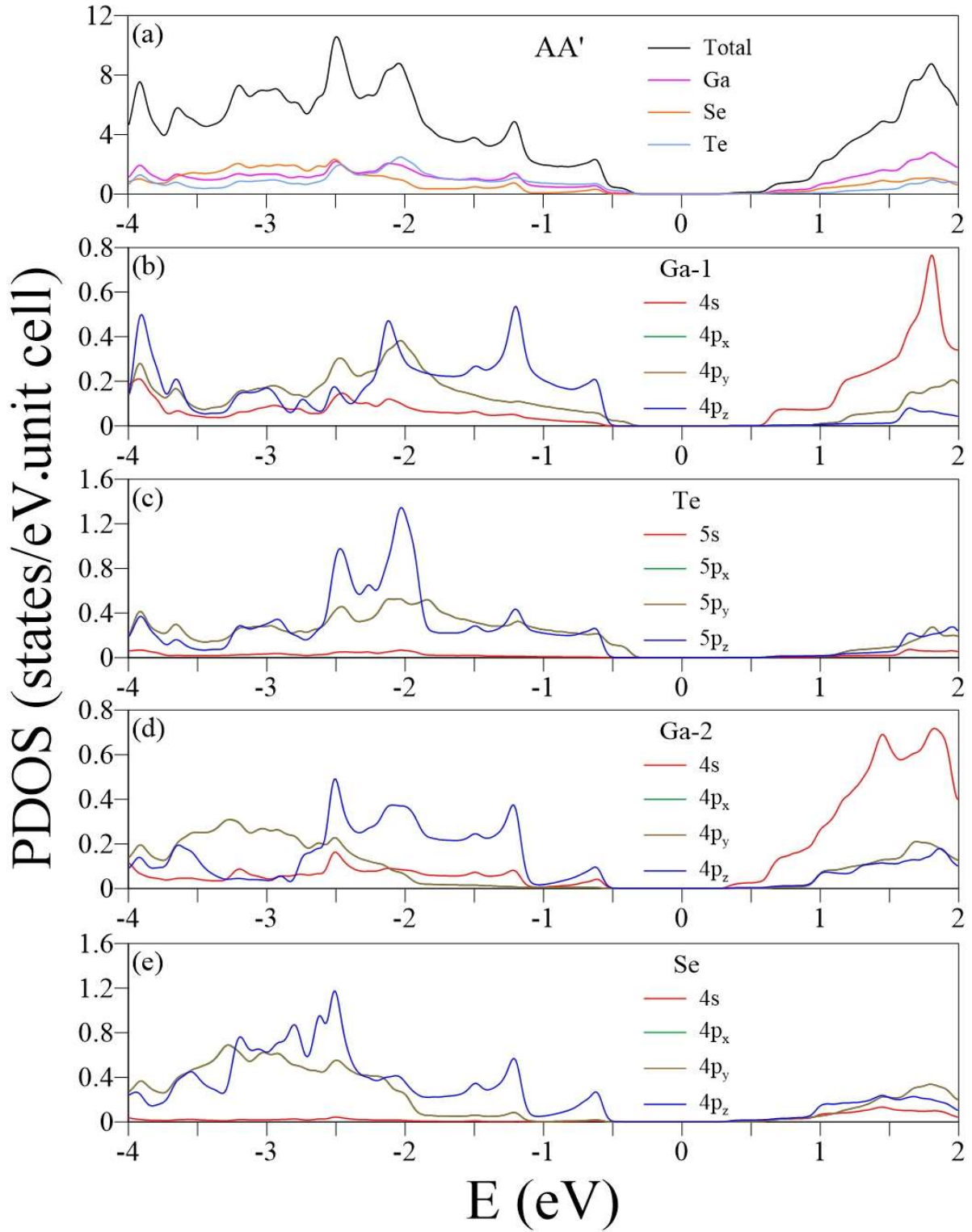

Figure 5: Projected density of states (PDOS) of the AA′-stacked GaSe/GaTe heterobilayer. (a) Total and atom-projected DOS for Ga, Se, and Te atoms. (b–e) Orbital-resolved PDOS for individual atoms: (b) Ga in the GaTe layer (Ga-1), (c) Te, (d) Ga in the GaSe layer (Ga-2), and (e) Se.

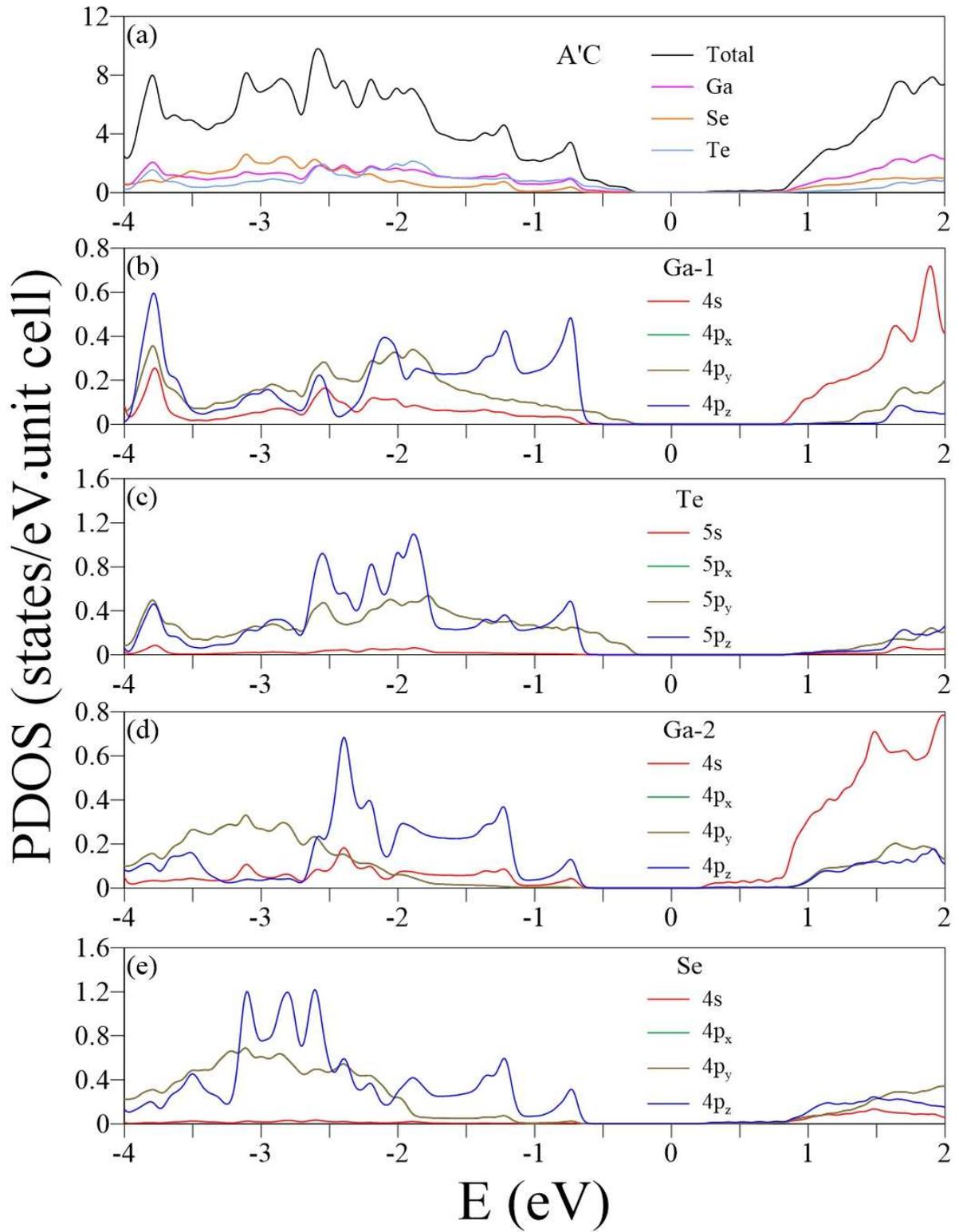

Figure 6: Projected density of states (PDOS) of the A′C-stacked GaSe/GaTe heterobilayer. (a) Total and atom-projected DOS for Ga, Se, and Te atoms. (b–e) Orbital-resolved PDOS for individual atoms: (b) Ga in the GaTe layer (Ga-1), (c) Te, (d) Ga in the GaSe layer (Ga-2), and (e) Se.

## Notes and references